\documentclass[american,aps,pra,reprint,superscriptaddress,twocolumn,showpacs]{revtex4-1}
\usepackage[unicode=true,pdfusetitle, bookmarks=true,bookmarksnumbered=false,bookmarksopen=false, breaklinks=false,pdfborder={0 0 0},backref=false,colorlinks=false] {hyperref}
\hypersetup{ colorlinks,linkcolor=myurlcolor,citecolor=myurlcolor,urlcolor=myurlcolor}
\usepackage{braket,colortbl,cleveref,amsthm,amsmath,amssymb,txfonts}
\definecolor{myurlcolor}{rgb}{0,0,0.7}
\usepackage{graphics,graphicx}
\usepackage{color}
\usepackage{graphicx}
\usepackage[format = plain,labelfont = bf,up, textfont = normal , up, justification =raggedright, singlelinecheck =false]{caption}
\usepackage{subcaption}

\theoremstyle{plain}

\def\bea{\begin{eqnarray}}
\def\eea{\end{eqnarray}}
\def\ba{\begin{array}}
\def\ea{\end{array}}

\def\ket{\rangle}
\def\bra{\langle}
\def\beq{\begin{equation}}
\def\eeq{\end{equation}}

\begin{document}

\title{Exact master equation for a spin interacting with a spin bath:\\
Non-Markovianity and negative entropy production rate}
\author{Samyadeb Bhattacharya}
\email{samyadebbhattacharya@hri.res.in}
\author{Avijit Misra}
\author{ Chiranjib Mukhopadhyay}
\author{Arun Kumar Pati}
\email{akpati@hri.res.in}
\affiliation{Harish-Chandra Research Institute, Allahabad 211019, India  and\\
 Homi Bhabha National Institute, Training School Complex, Anushakti Nagar, Mumbai 400 085, India}



\begin{abstract}
\noindent An exact canonical master equation of the Lindblad form is derived for a central spin interacting uniformly with a sea of completely unpolarized spins.  The Kraus operators for the dynamical map are also derived. The non-Markovianity of the dynamics in terms of the divisibility breaking of the dynamical map and increase of the trace distance fidelity between quantum states is shown. 
Moreover, it is observed that the irreversible entropy production rate is always negative (for a fixed initial state) whenever the dynamics exhibits non-Markovian behavior.
In continuation with the study of witnessing non-Markovianity, it is shown that the positive rate of change of the purity of the central qubit is a faithful indicator of the non-Markovian information back flow. Given the experimental feasibility of measuring the purity of a quantum state, a possibility of experimental demonstration of non-Markovianity and the negative irreversible entropy production rate is addressed. This gives the present work considerable practical importance for detecting the non-Markovianity and the negative irreversible entropy production rate.


\end{abstract}

\maketitle

\section{Introduction}
\label{I}
\noindent In many body problems 
the dynamics of microscopic (e.g. spin systems) or mesoscopic (e.g. SQUIDs) systems always gets complicated owing to its interaction with a background environment. To have the reduced dynamics of the quantum system that we are interested in, it is a general custom to model the environment as a collection of oscillators or spin half particles \cite{breuer} which is often abbreviated as bath. They constitute two different universal classes of quantum environment \cite{bath1}. In the oscillator bath model, the environment is described as a set of uncoupled harmonic oscillators. Paradigmatic examples of this kind of baths are spin-boson \cite{legget,weiss} and the Caldeira-Leggett model \cite{weiss,caldeira} originating from a scheme proposed by Feynman and Vernon \cite{feynman}. These oscillator models have been widely studied in the context of various physical phenomena
under Markovian approximation \cite{lindblad,gorini,breuer}.
On the other hand, the spin bath models remain relatively less explored. However, the spin bath models play a pivotal role in the quantum theory of magnetism \cite{magnets}, quantum spin glasses \cite{spinglass}, theory of conductors and superconductors \cite{conductor}. To get the exact dynamics of a quantum systems under this spin bath model is of paramount importance yet a difficult task. Indeed, in most of the cases the dynamics cannot be described exactly and several approximation techniques, both local and nonlocal in time, have been employed \cite{breuer,nakajima,
zwanzig,chaturvedi,breuer1,fischer,semin}.


 In this work, we will focus on the dynamical behavior of a central spin interacting uniformly with a spin bath and derive an exact time-local master equation of the Lindblad type. 
 Moreover, the Kraus representation of the dynamical map is also derived. Reduced dynamics of this particular spin bath model has been considered before \cite{fischer,semin} where correlated projection operator technique has been used to approximate the master equation of the central spin. However, the given master equation is time nonlocal and not of the standard canonical form.
In contrast, we start from the exact reduced state of the central spin at an arbitrary given time \cite{fischer} to derive the canonical master equation without considering any approximations. The thrust of our result is not only that the master equation is exact but the method used here allows us to unravel the less explored but far reaching consequences of the strong coupling regimes which can be instrumental in performing information theoretic, quantum thermodynamic and several other quantum technological tasks.  Moreover, the relaxation rates in the canonical master equation are insightful to understand several physical processes such as the dissipation, absorption and dephasing and thus the nature of decohrence. 

 One of the characteristics of the spin bath models is to exhibit the non-Markovian features \cite{breuer-review,huelga-review}. The non-Markovianity has been identified as a key resource in information theoretic \cite{info1,nment1,nment2}, thermodynamic \cite{nm-thermo,nm-thermo1,thermo-review} and precision measurement protocols \cite{metro1,metro2,metro3}. We study the non-Markovian features of the reduced dynamics and it is shown that the non-Markovianity increases with the interaction strength. 

 Irreversible increase of entropy due to dissipation of energy and work into the environment is inevitable for systems out of equilibrium. The analysis of irreversible or nonequilibrium entropy production and its rate have been instrumental to understand nonequilibrium phenomena in different branches of physics \cite{entropy1,entropy2,entropy3,entropy4,entropy5,entropy6}. According to the Spohn's theorem \cite{spohn}, the irreversible entropy production rate is always non-negative under the Markovian dynamics.  Whereas non-Markovianity of the dynamics allows negative irreversible entropy production rate and thereby this partial reversibility of the work and entropy influences the performance of quantum heat engines, refrigerators and memory devices. As our study enables us to probe the strong coupling regime, it can be far reaching to unravel the hitherto unexplored consequences of the non-Markovian dynamics in the strong coupling regime for more efficient thermodynamic protocols. Here, we 
 investigate the entropy production rate and shown that the non-Markovianity of the dynamics is always associated with a negative entropy production
rate of the central spin for a certain initial state. We also investigate the non-Markovianity in terms of the rate of change of the purity of the central qubit and it is observed that the rate of change of the purity  of the qubit is positive for the same aforesaid initial state, whenever the dynamics is non-Markovian. Experimental detection of the non-Markovianity and the entropy production rates for quantum systems are of paramount interest in current research. As purity can be measured in the laboratory, the study of this article can pave novel avenues to experimentally demonstrate non-Markovian features and negative entropy production rate in spin bath models.

 The organization of the paper is as follows. In Sec. \ref{II}, we derive the proposed canonical master equation of Lindblad type. The non-Markovian features of the dynamics of the central qubit are demonstrated explaining the indivisibility of the dynamical map and non-monotonicity of the trace distance fidelity. In this section, we also derive the Kraus operators for the dynamical evolution. The nonequilibrium entropy production rate and dynamics of purity of the qubit are studied in Sec. \ref{III}. Finally we conclude in Sec. \ref{IV}.

\section{Central spin model and its reduced dynamics}
\label{II}
In this section we first describe the central spin bath model. Then we derive the exact canonical master equation of the Lindblad type. From the master equation of the Lindblad form we show that the reduced dynamics of the central spin exhibits non-Markovian features throughout. We also derive the Kraus operators for the dynamical map. 

\subsection{The model}
 Let us first describe the central spin bath model. We consider a spin-$\frac{1}{2}$ particle that interacts uniformly with $N$ other spin-$\frac{1}{2}$ particles constituting the bath. The spins of the bath do not interact with each other. 
The Hamiltonian for this spin bath model is given by 
\begin{eqnarray}
 \label{sec1a}
H&=&H_s+H_{SB}\\ \nonumber
 &=&\frac{\hbar}{2}\omega_0\sigma_z^0+\frac{\hbar}{2}\sum_{i=1}^N \alpha(\sigma_x^0\sigma_x^i+\sigma_y^0\sigma_y^i+\sigma_z^0\sigma_z^i),
\end{eqnarray}
where $\sigma_k^i$ ($k=x,y,z$) are the Pauli matrices of the i-th spin of the bath and $\sigma_k^0$ ($k=x,y,z$) are the Pauli matrices for the central spin, $\alpha$ is the interaction strength. Here $H_S$ and $H_{SB}$ are the system and interaction Hamiltonian respectively.
Initially the system and reservoir is uncorrelated and the reservoir is in a thermal state at infinite temperature i.e., completely unpolarized state \cite{fischer}.
The composite state of the system and bath evolves unitarily under the total Hamiltonian $H +H_B$,  starting from the factorized initial state, $\rho_{SB}(0)=\rho_S(0)\otimes \frac{1}{2^N}\mathbb{I}_B$, where $\mathbb{I}_B$ is an $N$ qubit identity matrix and $H_B$ is the bath Hamiltonian. Note that as we are only concerned with the reduced dynamics of the central spin and the bath is completely unpolarized at $t=0$, there is no loss of generality to drop the bath Hamiltonian $H_B$ from the effective Hamiltonian $H$ to get the reduced dynamics of the spin.
%
Therefore, the reduced quantum state $\rho_{S}(t)$  of the central spin at time $t$,  can be obtained by tracing out the bath degrees of freedom as
\beq
\label{u-evolution}
\rho_{S}(t)=\mbox{Tr}_B[e^{-i(H+H_B)t/\hbar}\{\rho_S(0)\otimes \frac{1}{2^N}\mathbb{I}_B\}e^{i(H+H_B)t/\hbar}].
\eeq
Hereafter, we drop the subscript $_S$ for brevity to denote central spin as we will only deal with it.
The total angular momentum of the bath is given by $
\mathbf{J}= \frac{1}{2} \sum_i\boldsymbol{\sigma}^i.$
 The basis $|j,m\ket$ is defined as the simultaneous eigenbases of both $\mathbf{J}^2$ and $J_z$. For even $N$, $j$ takes the values $j=0,1,2,....N/2$  and for odd N, we have $j=1/2,3/2,...N/2$ and $m$ goes from $-j$ to $j$.  
It can be shown that \cite{fischer} the \textit{z}-component of the  total angular momentum $\frac{1}{2} \sigma_{z}^{0} +  \frac{1}{2} \sum_i\ \sigma_z^i $  as well as $\boldsymbol{J}^{2}$ are conserved quantities. There are now two dimensional subspaces spanned by $|+\rangle \otimes |j,m \rangle $ and $ |- \rangle \otimes |j,m+1 \rangle $ which are invariant under time evolution. Now the task of finding the analytical solution to the reduced dynamics of the central spin is  broken down into solving the equations of motion in each subspace. 
Solving the equation of motion exactly\cite{fischer}, the initial reduced state of the central spin, $\rho= \left(\begin{matrix}\rho_{11} && \rho_{12}\\\rho_{21} && \rho_{22}\end{matrix}\right)$ can be shown to evolve  as  
\begin{eqnarray}
\label{dynamics}
 \rho_{11}(t)&=&A(t)\rho_{11}(0)+B(t)\rho_{22}(0),\nonumber \\
 \rho_{12}(t)&=& C(t)\rho_{12}(0).
\end{eqnarray}
Where,
\beq\label{sec1pN}
\begin{array}{ll}
A(t)=\sum_{j,m}\frac{N_j}{2^N} \left[\cos^2\left(\mu_+(j,m)t\right)+\frac{\Omega_+^2(m)}{4\mu_+^2(j,m)}\sin^2\left(\mu_+(j,m)t\right)\right],\nonumber\\
B(t)=\sum_{j,m}\frac{N_j}{2^N}\frac{\alpha^2 b^2(j,m)}{4\mu_+^2(j,m)}\sin^2\left(\mu_+(j,m)t\right), \nonumber\\
C(t)=e^{i\omega_0t}\sum_{j,m}\frac{N_j}{2^N}\left[\cos\left(\mu_+(j,m)t\right)-\frac{i\Omega_+(m)}{2\mu_+(j,m)}\sin\left(\mu_+(j,m)t\right)\right]\\
~~~~~~~~~~~~~~~~~~\times\left[\cos\left(\mu_-(j,m)t\right)+\frac{i\Omega_-(m)}{2\mu_-(j,m)}\sin\left(\mu_-(j,m)t\right)\right],
\end{array}
\eeq 
and
$
\begin{array}{ll}
N_j=\left(\begin{matrix}   
          N\\
          \frac{N}{2}+j     
     \end{matrix}\right)
     -\left(\begin{matrix}   
          N\\
          \frac{N}{2}+j+1     
     \end{matrix}\right),\\
     \\
\begin{array}{ll}
\Omega_{\pm}=\pm\omega_0+\alpha(\pm m +1/2),\\
\mu_{\pm}=\frac{1}{2}\sqrt{\Omega_{\pm}^2+\alpha^2b_{\pm}^2},\\
b_{\pm}=\sqrt{j(j+1)-m(m \pm 1).}
\end{array}
\end{array}
$

It follows from the above expressions that $A(t)+B(t)=1$, which implies the dynamical map is unital. The unitality of the dynamics has to be satisfied as the environment and the systems starts from a product state while the environment being in the maximally mixed state. We are now in position to derive the canonical master equation.
\subsection{Canonical master equation} 
Derivation of the master equation is basically finding the generator of the evolution, which is one of the fundamental problems in the theory of open quantum systems. Moreover, the Lindblad type master equation can lead to understanding of various physical processes like dissipation, absorption, dephasing and hence the nature of decoherence, in a much more convincing way. Considering the importance of the spin bath to model the environmental interactions in various domains, such as magnetism, superconductors, spin glasses etc., it is of  important and illustrative to have the master equation for spin bath models. Additionally, theoretical as well as experimental study of quantum thermodynamic devices (QTDs) has attracted a great deal of interest in recent times. Establishing master equations for open quantum systems is of paramount significance in the context of QTDs \cite{kurizki-review}, where a single or  few quantum systems are coupled with their heat baths in general. For example, in recently proposed  
quantum absorption refrigerators \cite{
popescu-refri}, three qubits interact among themselves while they are coupled to their respective baths. The Lindblad operators for the qubits under the corresponding heat baths become crucial to the study the performance of the refrigerators in both steady and transient regimes \cite{kosloff-refri, huber-cavity, huber-josephson, correa-lindblad, huber-transient,sreetama}. Recently introduced quantum thermal transistors \cite{transistor} are also worth mentioning in this context. Therefore, the canonical Lindblad-type master equation in the spin bath models can provide a novel way to study the QTDs in hithertho less explored strong coupling and non-Markovian regime which might have far-reaching impacts to enhance the performance of QTDs.

In what follows, we derive the exact canonical master equation of the Lindblad type for the central spin starting from dynamical map given in Eq. \eqref{dynamics}. The dynamical map described in Eq. \eqref{dynamics} can be notationally represented as 
\beq\label{sec1f}
\rho(t)=\Phi[\rho(0)].
\eeq
 The equation of motion of the reduced density matrix of the form
\beq\label{sec1g}
\dot{\rho}(t)=\Lambda[\rho(t)]
\eeq
can be obtained from Eq. \eqref{dynamics}, which is characterized by the time dependent generator $\Lambda[.]$.  
By following the method \citep{andersson} given below, we find the master equation and thus the generator of the specific reduced dynamics.
Consider a convenient orthonormal basis set $\{G_a\}$ with the properties $G_a^{\dagger}=G_a$ and $Tr[G_aG_b]=\delta_{ab}.$
%
%
%
The map given in Eq. \eqref{sec1f} can now be represented as 
\beq\label{sec1j}
\Phi[\rho(0)]=\sum_{k,l} \mbox{Tr}[G_k\Phi[G_l]]\mbox{Tr}[G_l\rho(0)]G_k = [F(t)r(0)]G^T,
\eeq
 where $F_{kl} = \mbox{Tr}[G_k\Phi[G_l]],~~~~r_l = \mbox{Tr}[G_l\rho(0)]$.
 Differentiating Eq. \eqref{sec1j}, we get 
\beq\label{sec1l}
\dot{\rho}(t) = [\dot{F}(t)r(0)]G^T.
\eeq
Let us consider a matrix $L$, with elements $L_{kl}=\mbox{Tr}[G_k\Lambda[G_l]].$
%
%
 We can now represent Eq. \eqref{sec1g} as 
\beq\label{sec1n}
\dot{\rho}(t)=\sum_{k,l} \mbox{Tr}[G_k\Lambda[G_l]]\mbox{Tr}[G_l\rho(t)]G_k = [L(t)r(t)]G^T.
\eeq
By comparing Eq. \eqref{sec1l} and \eqref{sec1n}, we find 
\beq\label{sec1o}
\dot{F}(t)=L(t)F(t)~\Rightarrow L(t)=\dot{F}(t)F(t)^{-1}.
\eeq
 We can arrive at Eq. \eqref{sec1o} given the inverse of $F(t)$ does exist and $F(0)=\mathbb{I}$.
Considering the specific map of the central spin in Eq. \eqref{dynamics},  
and taking the Orthonormal basis set $\lbrace G_{a} \rbrace$ as $\lbrace \frac{\mathbb{I}_2}{\sqrt{2}},\frac{\sigma_{x}}{\sqrt{2}},\frac{\sigma_{y}}{\sqrt{2}},\frac{\sigma_{z}}{\sqrt{2}} \rbrace$, we find the $L(t)$ matrix to be 
\begin{widetext}
\beq\label{sec1t}
L(t)=\left(\begin{matrix}
0 & 0 & 0 & 0\\
0 & \frac{C_R(t)\dot{C}_R(t)+C_I(t)\dot{C}_I(t)}{C_R(t)^2+C_I(t)^2} & -\frac{C_I(t)\dot{C}_R(t)-C_R(t)\dot{C}_I(t)}{C_R(t)^2+C_I(t)^2} & 0\\
0 & \frac{C_I(t)\dot{C}_R(t)-C_R(t)\dot{C}_I(t)}{C_R(t)^2+C_I(t)^2} & \frac{C_R(t)\dot{C}_R(t)+C_I(t)\dot{C}_I(t)}{C_R(t)^2+C_I(t)^2} & 0\\
(\dot{A}(t)+\dot{B}(t))+\left( \frac{\dot{A}(t)-\dot{B}(t)}{A(t)-B(t)} \right)(1-(A(t)+B(t))) & 0 & 0 & \left( \frac{\dot{A}(t)-\dot{B}(t)}{A(t)-B(t)} \right) \\
\end{matrix}\right),
\eeq
\end{widetext}
where $C_R(t)$ and $C_I(t)$ are the real and imaginary part of $C(t)$ respectively. Now from the Eq. \eqref{sec1n}, we get the equation of motion as given by
%
%
\beq\label{sec1v}
\begin{array}{ll}
\dot{\rho}_{11}(t)=\frac{L_{z0}+L_{zz}}{2}\rho_{11}(t)+\frac{L_{z0}-L_{zz}}{2}\rho_{22}(t),\\
\dot{\rho}_{12}(t)=(L_{xx}+iL_{xy})\rho_{12}(t).
\end{array}
\eeq
Eq. \eqref{sec1v} gives the time rate of change of the density matrix. However, one needs to have the Lindblad type master equation to understand various processes like dissipation, absorption, dephasing in a more convincing way. Moreover, it is of prime importance to have the master equation to study the non-Markovian behavior of the reduced dynamics as we will see later. Therefore, our immediate aim is to derive the Lindblad type master equation starting from Eq. \eqref{sec1v}. Eq. \eqref{sec1g} can be written in the form \cite{hall}
\beq\label{sec1A}
\dot{\rho}(t)=\Lambda[\rho(t)]=\sum_{k} X_k(t) \rho(t)Y_{k}(t)^{\dagger},
\eeq
where $X_k(t)=\sum_i G_i x_{ik}(t)$,$Y_k(t)=\sum_i G_i y_{ik}(t)$  and $\{G_a\}$ are the basis operators as defined before.
Using this decomposition of $X(t)$ and $Y(t)$, Eq. \eqref{sec1A} can be rewritten as 
\beq\label{sec1C}
\dot{\rho}(t)= \sum_{i,j=\{0,x,y,z\}} z_{ij}(t)G_i\rho(t)G_j,
\eeq 
\noindent where $z_{ij}(t)=\sum_{k} x_{ik}(t)y_{jk}(t)^{*}$ are the elements of a Hermitian matrix.
%
Using a new set of operators \cite{hall} $\mathcal{F}(t)=(z_{00}(t)/8)\mathbb{I}_2+\sum_{i}(z_{i0}/2)G_i$ and $H(t)=\frac{i}{2}\hbar(\mathcal{F}(t)-\mathcal{F}^{\dagger}(t))$, after some algebra,  the Eq. \eqref{sec1C} can be written as
%
%
\beq\label{sec1F}
\begin{array}{ll}
\dot{\rho}(t)=\frac{i}{\hbar}[\rho(t),H(t)]\\
+\sum_{i,j=\{x,y,z\}} z_{ij}(t)\left(G_i\rho G_j-\frac{1}{2}\{G_jG_i,\rho(t)\}\right),
\end{array}
\eeq
where the curly braces stand for anti-commutator. Hence, the canonical master equation of the Lindblad form read as
\beq\label{sec1G}
\begin{array}{ll}
\dot{\rho}(t)=\frac{i}{\hbar}U(t)[\rho(t),\sigma_z]+\Gamma_{deph}(t)\left[\sigma_z\rho(t)\sigma_z-\rho(t)\right]\\
~~~~~~+\Gamma_{dis}(t)\left[\sigma_{-}\rho(t)\sigma_{+}-\frac{1}{2}\{\sigma_{+}\sigma_{-},\rho(t)\}\right]\\~~~~~~+\Gamma_{abs}(t)\left[\sigma_{+}\rho(t)\sigma_{-}-\frac{1}{2}\{\sigma_{-}\sigma_{+},\rho(t)\}\right],
\end{array}
\eeq
where $\sigma_{\pm}=\frac{\sigma_x \pm i\sigma_y}{2}$, and $\Gamma_{dis}(t), \Gamma_{abs}(t), \Gamma_{deph}(t)$ are the rates of dissipation, absorption and dephasing processes respectively , and  $U(t)$ corresponds to the unitary evolution. $A(t)+B(t)=1$, for this specific system, is used to derive the master equation. 
The rates of dissipation, absorption, dephasing and the unitary evolution are, respectively, given as
\beq\label{sec1I}
\begin{array}{ll}
\Gamma_{dis}(t)=-\frac{L_{z0}+L_{zz}}{2}=\frac{d}{dt}\left[\ln\left(\frac{1}{\sqrt{A(t)-B(t)}}\right)\right],\\
\\
\Gamma_{abs}(t)=-\frac{L_{zz}-L_{z0}}{2}=\frac{d}{dt}\left[\ln\left(\frac{1}{\sqrt{A(t)-B(t)}}\right)\right],\\
\\
\Gamma_{deph}(t)=-\frac{2L_{xx}-L_{zz}}{4}=\frac{1}{4}\frac{d}{dt}\left[\ln\left(\frac{A(t)-B(t)}{|C(t)|^2}\right)\right],\\
\\
U(t)=-\frac{L_{xy}}{2}=-\frac{1}{2}\frac{d}{dt}\left[\ln\left(1+\left(\frac{C_R(t)}{C_I(t)}\right)^2\right)\right].
\end{array}
\eeq
Note that the system environment interaction generates a time dependent driving Hamiltonian evolution in the form of $U(t)$. 
Since the coefficients of dissipation and absorption are equal, the master equation (\ref{sec1G}) can also be rewritten as 
\beq\label{sec1Ia}
\begin{array}{ll}
\dot{\rho}(t)=\frac{i}{\hbar}U(t)[\rho(t),\sigma_z]
+\frac{\Gamma_{dis}(t)}{2}\left[\sigma_x\rho(t)\sigma_x-\rho(t)\right]\\
+\frac{\Gamma_{dis}(t)}{2}\left[\sigma_y\rho(t)\sigma_y-\rho(t)\right]
+\Gamma_{deph}(t)\left[\sigma_z\rho(t)\sigma_z-\rho(t)\right].
\end{array}
\eeq
The above equation implies that $\frac{\mathbb{I}_2}{2}$ is a fixed point of the reduced dynamics  and hence, it confirms the unitality of the dynamical map.
 As the bath is in a thermal state at infinite temperature, the probabilities of losing energy to the bath modes and absorbing from it become equal which makes the dissipation and absorption rates to be the same. This is quite similar to the bosonic thermal baths, as it follows from the KMS condition \cite{breuer} that given the baths having canonical equilibrium distribution the rates of the absorption and dissipation process are balanced by the equation $\Gamma(-\omega)=\Gamma(\omega)\exp(-\beta\omega)$. Here $\beta$ is the inverse temperature of the bath and it implies that $\Gamma(-\omega)=\Gamma(\omega)$, iff $\beta=0$.

One of the important properties of a quantum dynamical map is completely positivity \cite{breuer-review,huelga-review,akr1,akr2,wolf,breuer2,laine,rivas,raja,breuer3}. 
The notion ``complete" comes with the argument that for any valid quantum dynamical map, the positivity must be preserved if the map is acting on a system which is correlated to an ancilla of any possible dimension. For a Lindblad type canonical master equation with time dependent coefficients, as in Eq. \eqref{sec1G}, the complete positivity is guaranteed by the following condition $\int_0^t \Gamma_i(s)ds \geq 0$ \cite{kossa}, which can be easily verified for the specific decay rates given in (\ref{sec1I}). It is worth mentioning that since the dynamical map for this specific spin bath model is derived starting from an initial product system plus environment state, it is always guaranteed to be completely positive \cite{alicki,pechukas}. However, the complete positivity of the dynamical map for the reduced system can break down in the presence of system-environment initial correlation \cite{pechukas}.
\subsection{Operator sum representation}
Other important aspect of general quantum evolution is the Kraus operator sum representation, given as $\rho(t)=\sum_i K_i(t)\rho(0)K_i^{\dagger}(t)$. The Kraus operators can be constructed \cite{leung} from the eigenvalues and eigenvectors of the corresponding Choi-Jamiolkowski state \cite{choicp}.
The Choi-Jamiolkowski state for a dynamical map $\Phi[\rho]$ acting on a $d$ dimensional system is given by $(\mathbb{I}_d \otimes\Phi)[\Phi_{+}]$, with $\Phi_{+}=|\Phi_{+}\rangle\langle\Phi_{+}|$ being the maximally entangled state in $d\times d$ dimension. For the particular evolution considered here, we find the Choi state to be 
\beq\label{sec1newb}
\left(
\begin{matrix}
A(t)/2 && 0 && 0 && C(t)/2\\
0 && B(t)/2 && 0 && 0\\
0 && 0 && B(t)/2 && 0\\
C^*(t)/2 && 0 && 0 && A(t)/2
\end{matrix}
\right).
\eeq
The positive semi-definiteness of the above density matrix demands $B(t)\geq 0~;~A(t) \geq |C(t)|$.
From the eigensystem of the Choi state given in (\ref{sec1newb}), we derive the Kraus operators as  
\beq\label{sec1newa}
\begin{array}{ll}
K_1(t)=\sqrt{B(t)}\left(\begin{matrix}
                        0 && 1\\
                        0 && 0
                        \end{matrix}\right),\\
\\
K_2(t)=\sqrt{B(t)}\left(\begin{matrix}
                        0 && 0\\
                        1 && 0
                        \end{matrix}\right),\\
\\
K_3(t)=\sqrt{\frac{A(t)-|C(t)|}{2}}\left(\begin{matrix}
                        -e^{i\theta(t)} && 0\\
                        0 && 1
                        \end{matrix}\right),\\ 
\\
K_4(t)=\sqrt{\frac{A(t)+|C(t)|}{2}}\left(\begin{matrix}
                        e^{i\theta(t)} && 0\\
                        0 && 1
                        \end{matrix}\right),                       
\end{array}
\eeq
where $\theta(t)=\arctan[C_I(t)/C_R(t)]$. It is straight forward to verify that the Kraus operators satisfies the unitality property  $ \sum_i K_i(t) K_i^{\dagger}(t)=\mathbb{I}$. 
\subsection{Non-Markovianity}
The charecterization and quantification of the non-Markovianity is a fundamental aspect of open quantum dynamics. There are several proposed measures based on CP divisibility \cite{rivas,kossa1} and non-Markovianity witness \cite{laine,breuer2,vasile,lu,luo,fanchini,titas,salimi}. 
One of the well accepted characterization and quantification of non-Markovianity based on the composition law of the dynamical map has been introduced by Rivas-Huelga-Plenio \cite{rivas}, commonly known as RHP measure of non-Markovianity.
 In this approach, the non-Markovian behaviour is attributed to the deviation from divisibility and the quantification of non-Markovianity is done based on the amount of the deviation. A complete positive and trace preserving (CPTP) dynamical map $\Phi(t,0)$ is divisible, when for all intermediate time $\tau$, it follows that
\beq\label{sec1O}
\Phi(t+\tau,0)=\Phi(t+\tau,t)\Phi(t,0).
\eeq
In Ref. \cite{rivas}, it has been shown that the dynamical map $\Phi$ is divisible or indivisible if the (right) time derivative
\beq\label{sec1P}
q(t)=\lim_{\epsilon\rightarrow 0^+} \frac{||\left(\mathbb{I}_d\otimes\Phi(t+\epsilon,t)\right)\Phi_+||-1}{\epsilon},
\eeq
is zero or greater than zero, respectively. Here, $d$ is the dimension of the Hilbert space and $||.||$ denotes for trace norm  and $\Phi_{+}=|\Phi_{+}\rangle\langle\Phi_{+}|$ is the maximally entangled state in $d\times d$ dimension. To illustrate this measure, we consider the dynamical equation (\ref{sec1g}). In the limit $\epsilon\rightarrow 0^+$, The solution formally reads $\Phi(t+\epsilon,t)\rightarrow e^{\epsilon\Lambda}$. To the first order expansion, the parameter $q(t)$ is given as
\beq\label{sec1Pnew}
q(t)=\lim_{\epsilon\rightarrow 0^+} \frac{||\left(\mathbb{I}_{d\times d}+\mathbb{I}_d\otimes\Lambda(t+\epsilon,t)\right)\Phi_+||-1}{\epsilon},
\eeq
It is strightforward to calculate $q(t)$ from Eq. (\ref{sec1Pnew}) \cite{rivas}. Hence, the RHP measure of non-Markovianity can be defined \cite{kossa1,rivas} based on the strict positivity of $q(t)$ as follows
\beq\label{sec1Q}
G = \frac{\eta}{\eta+1},
\eeq
where $\eta=\int_0^{\infty}q(t)dt$. Note that for the Markovian evolution $G$ is zero and the maximum non-Markovianity corresponds to $G=1$, i.e., when $\eta \rightarrow \infty$. The positivity of the function $q(t)$ or indivisibility of the map appears when the relaxation rates ($\{\Gamma_i(t)\}$s) take negative values. We will show in the following that for the specific dynamical evolution considered in the present work, the decay rates periodically get negative and hence break the divisibility of the map, although always maintain the complete positivity condition. For this particular evolution, we get
\beq\label{sec1R}
\begin{array}{ll}
q(t)=[|\Gamma_{dis}(t)|-\Gamma_{dis}(t)] + [|\Gamma_{deph}(t)|-\Gamma_{deph}(t)]\\
~~~~~~=q_{dis}(t)+q_{deph}(t),
\end{array}
\eeq
where $q_{deph}(t)=|\Gamma_{deph}(t)|-\Gamma_{deph}(t)$, is the non-Markovianity for the dephasing channel and $q_{dis}(t)=|\Gamma_{dis}(t)|-\Gamma_{dis}(t)$, is that for the thermal part of the channel including the dissipation and absorption process.

\begin{figure}[htb]
\includegraphics[width=7cm, height=6cm]{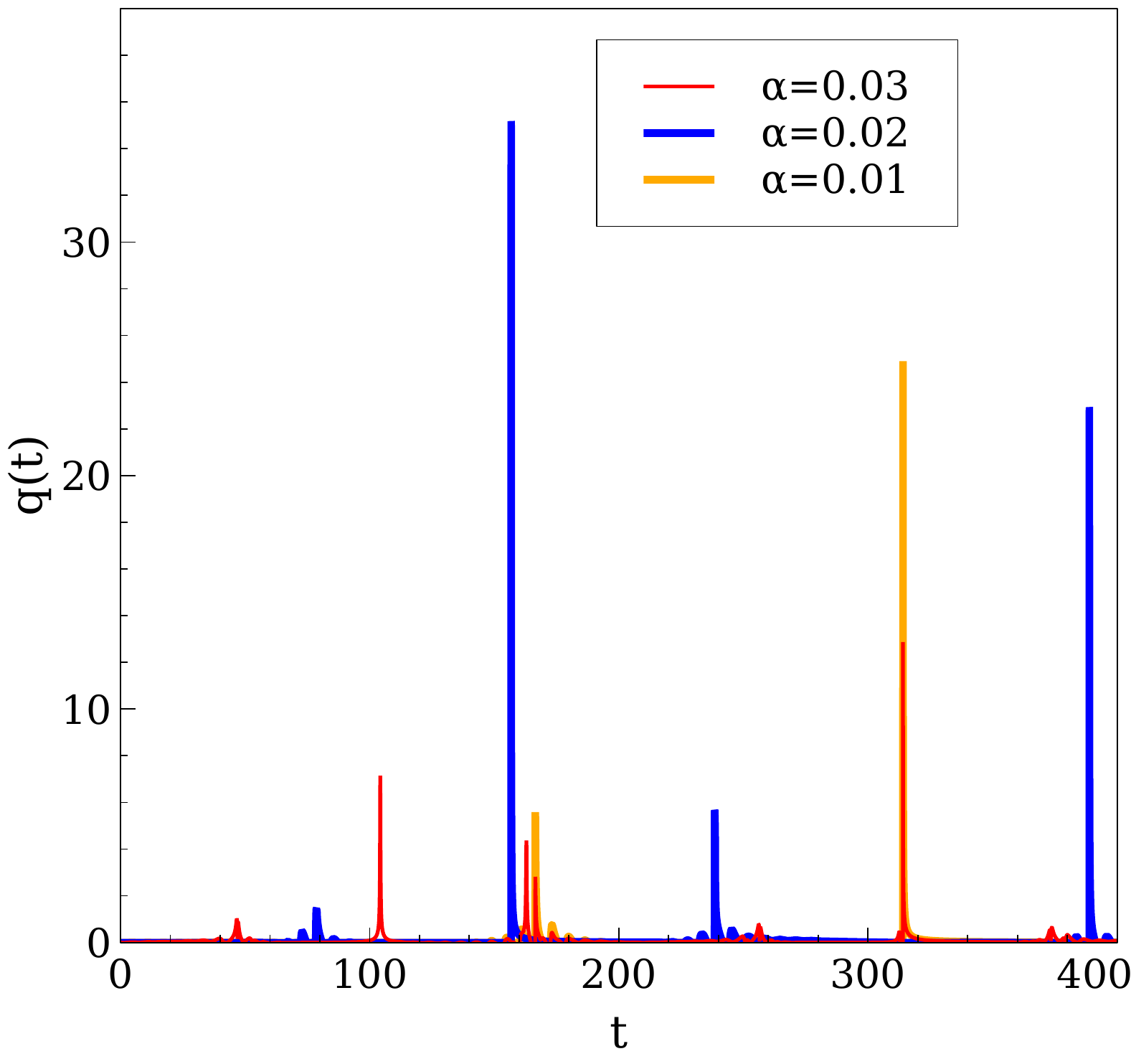}
\caption{(Colour online) Variation of $q(t)$ with time $t$ for various interaction strength $\alpha$. Number of bath spins is kept fixed at $N=20$. Positive $q(t)$ implies the non-Markovian nature of the dynamics according to the RHP measure.  }
\label{NM-RHP-fig1}
\end{figure}
\begin{figure}[htb]
                \includegraphics[width=7cm, height=6cm]{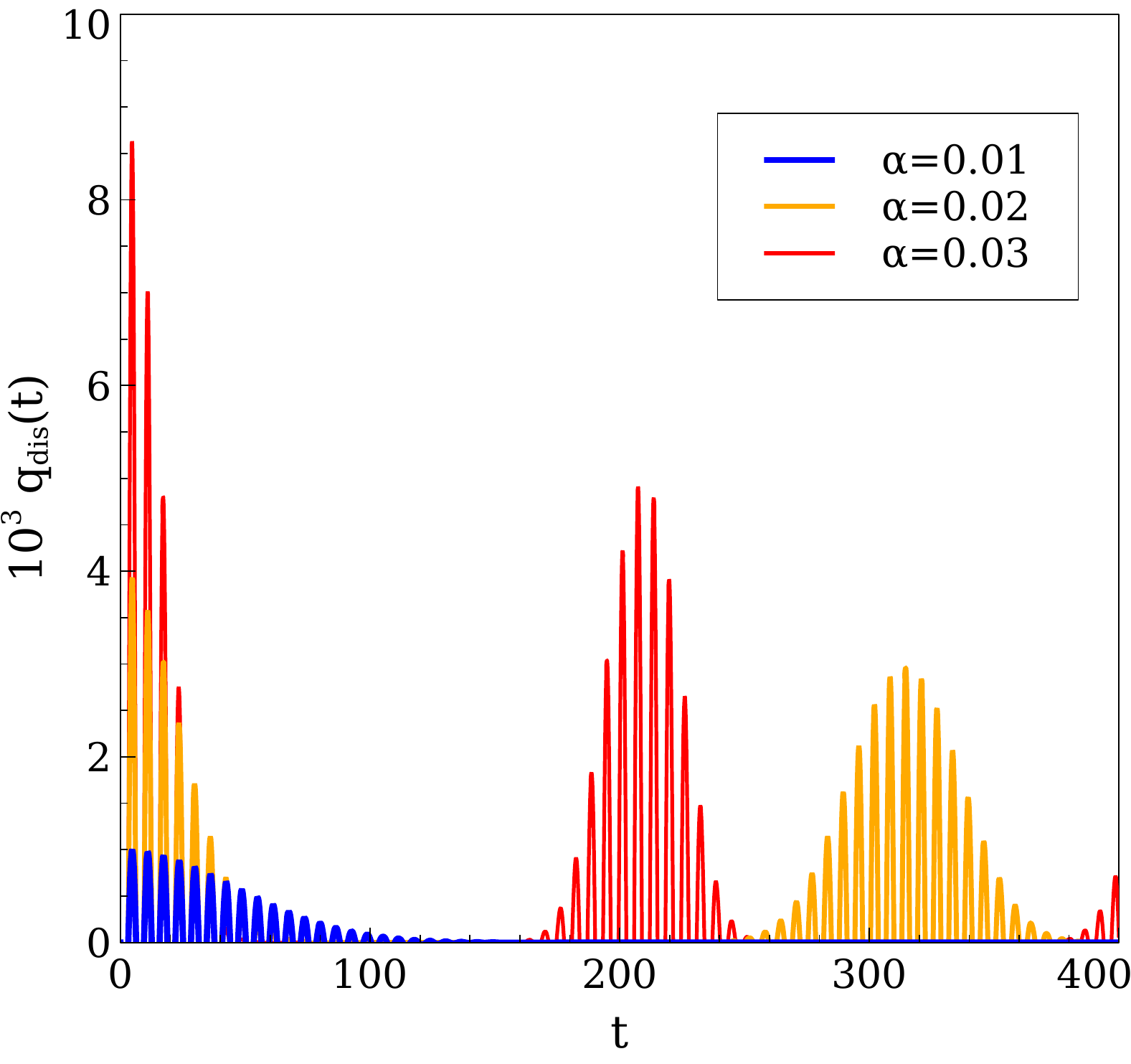}
               \caption{(Colour online) Variation of $q_{dis}(t)$ with time $t$ for various interaction strength $\alpha$. Number of bath spins is kept fixed at $N=20$. To distinguish the effect on the thermal part of the quantum channel, we separately plot $q_{dis}(t)$. It can be seen from the plot that the non-Markovian revival for the thermal part of the channel increases with the increase of the interaction strength $\alpha$ for fixed $N$.}
                \label{NM-RHP-fig2}
\end{figure}

\begin{figure}[htb]
\includegraphics[width=7cm, height=6cm] {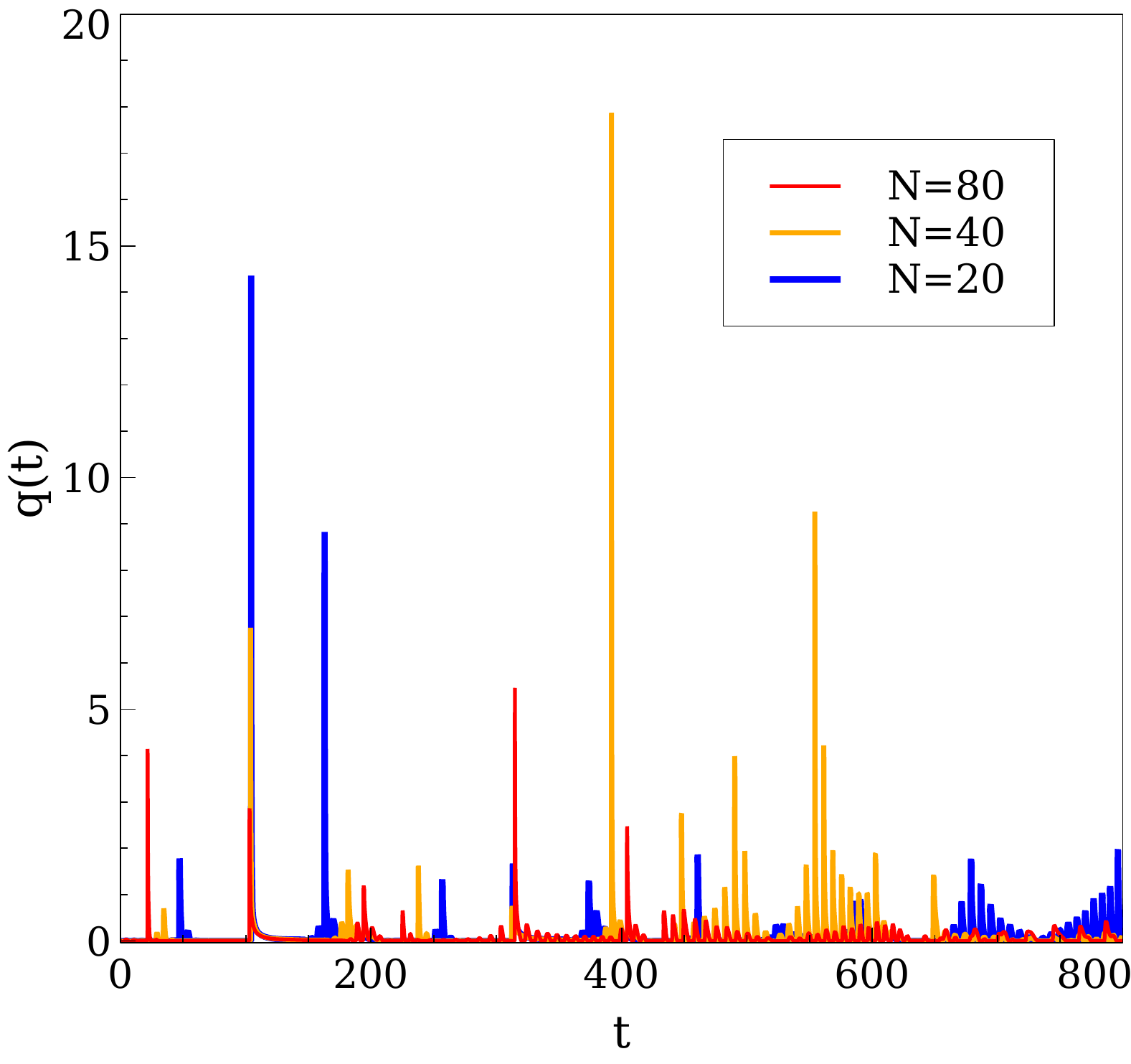}
	\caption{(Colour online)  Variation of $q(t)$ with time $t$ for different number of bath spins $N$. Interaction strength $\alpha=0.03$ is taken. }
	\label{NM-RHP-N} 
\end{figure}

\begin{figure}[htb]
\includegraphics[width=7cm, height=6cm] {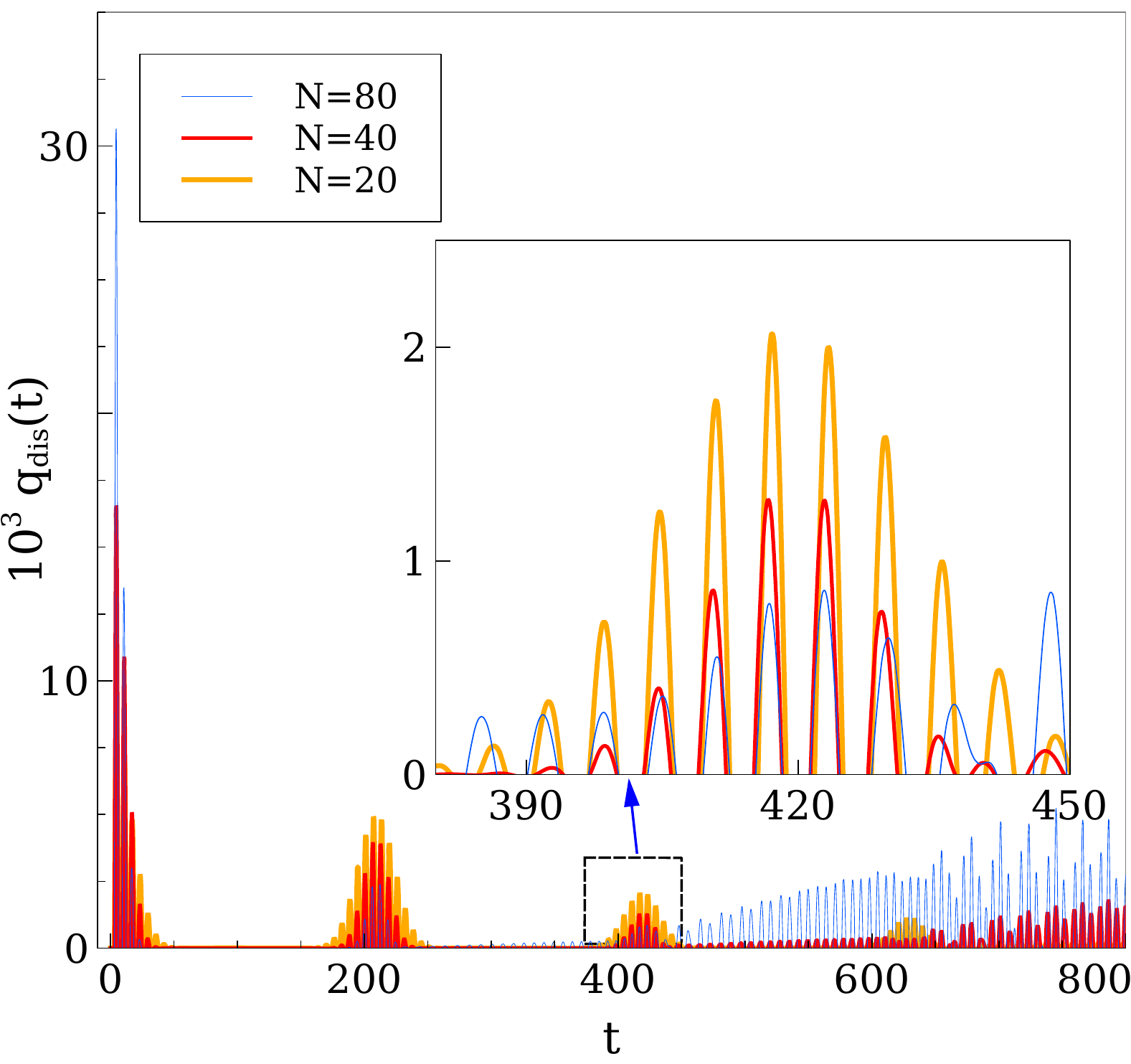}
	\caption{(Colour online)  Variation of $q_{dis}(t)$ with time $t$ for different number of bath spins $N$. Interaction strength $\alpha=0.03$ is taken. Magnified view of the rectangular region is shown in the inset. The plot depicts that the revival of $q_{dis}(t)$ increases with the increase of bath spins $N$.}
	\label{NM-RHP-N1} 
\end{figure}

In Fig. \ref{NM-RHP-fig1} and \ref{NM-RHP-fig2}, we plot the total non-Markovianity $q(t)$ and the contribution due to the thermal channel $q_{dis}(t)$ with different values of $\alpha$ to show the non-Markovian behavior of the dynamics. We see that the revival of $q_{dis}(t)$ increases with the increasing interaction strength $\alpha$. In Fig. \ref{NM-RHP-N} and \ref{NM-RHP-N1}, we plot $q(t)$ and total non-Markovianity $q_{dis}(t)$ respectively, but for different number of bath spins $N$ with a fixed interaction strength.

\begin{figure}[htb]
	\includegraphics[width=7cm, height=6cm] {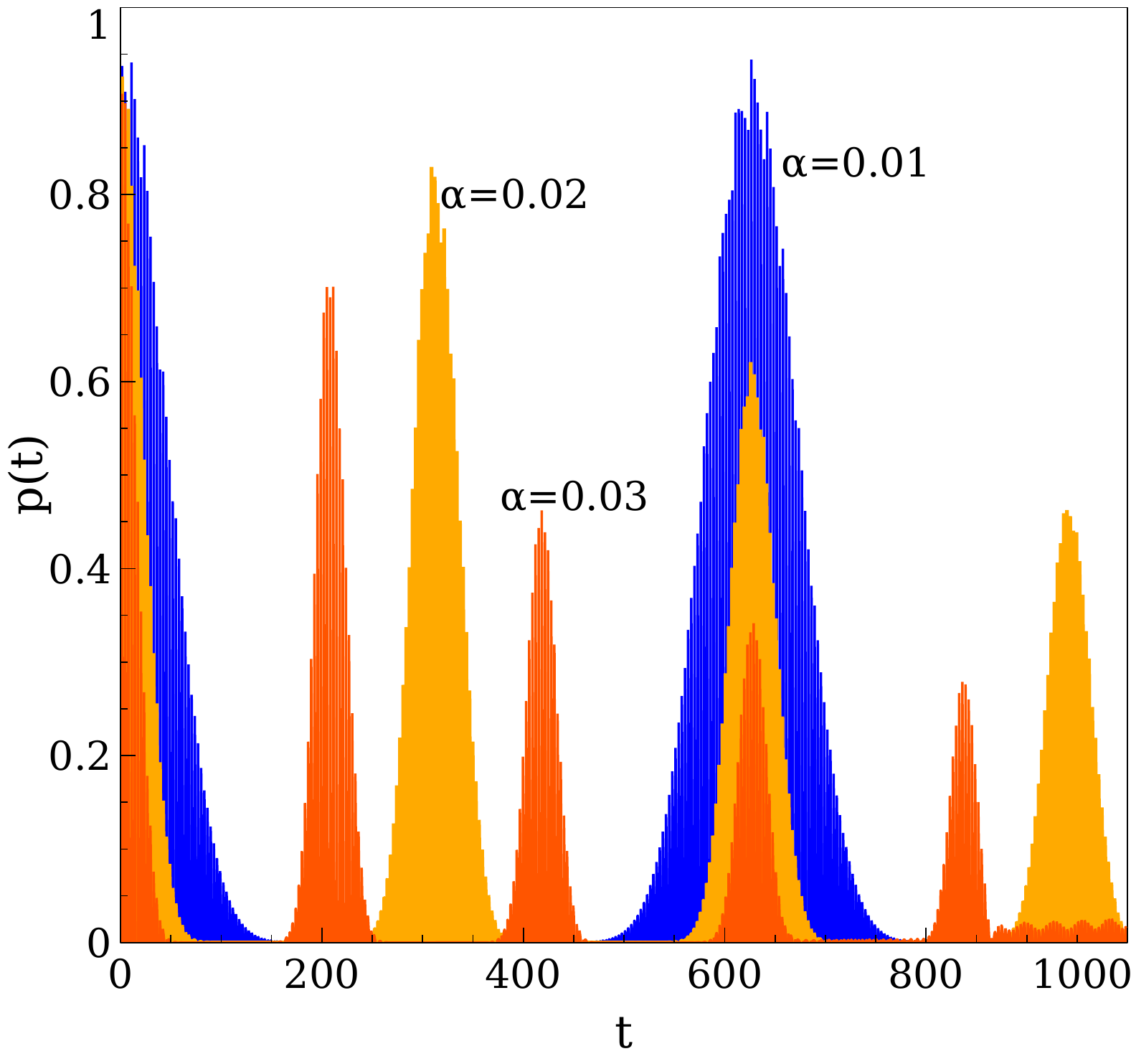}
	\caption{(Colour online) Variation of $p(t)$ with time $t$ for the two states $|\pm\ket=\frac{1}{\sqrt{2}}|0 \pm 1\ket$, for different interaction strength $\alpha$ (where $N=20$). Positive $p(t)$ implies non-Markovianity according to the BLP measure.}
	\label{blp-fig}
\end{figure}

\begin{figure}[htb]
	\includegraphics[width=7cm, height=6cm] {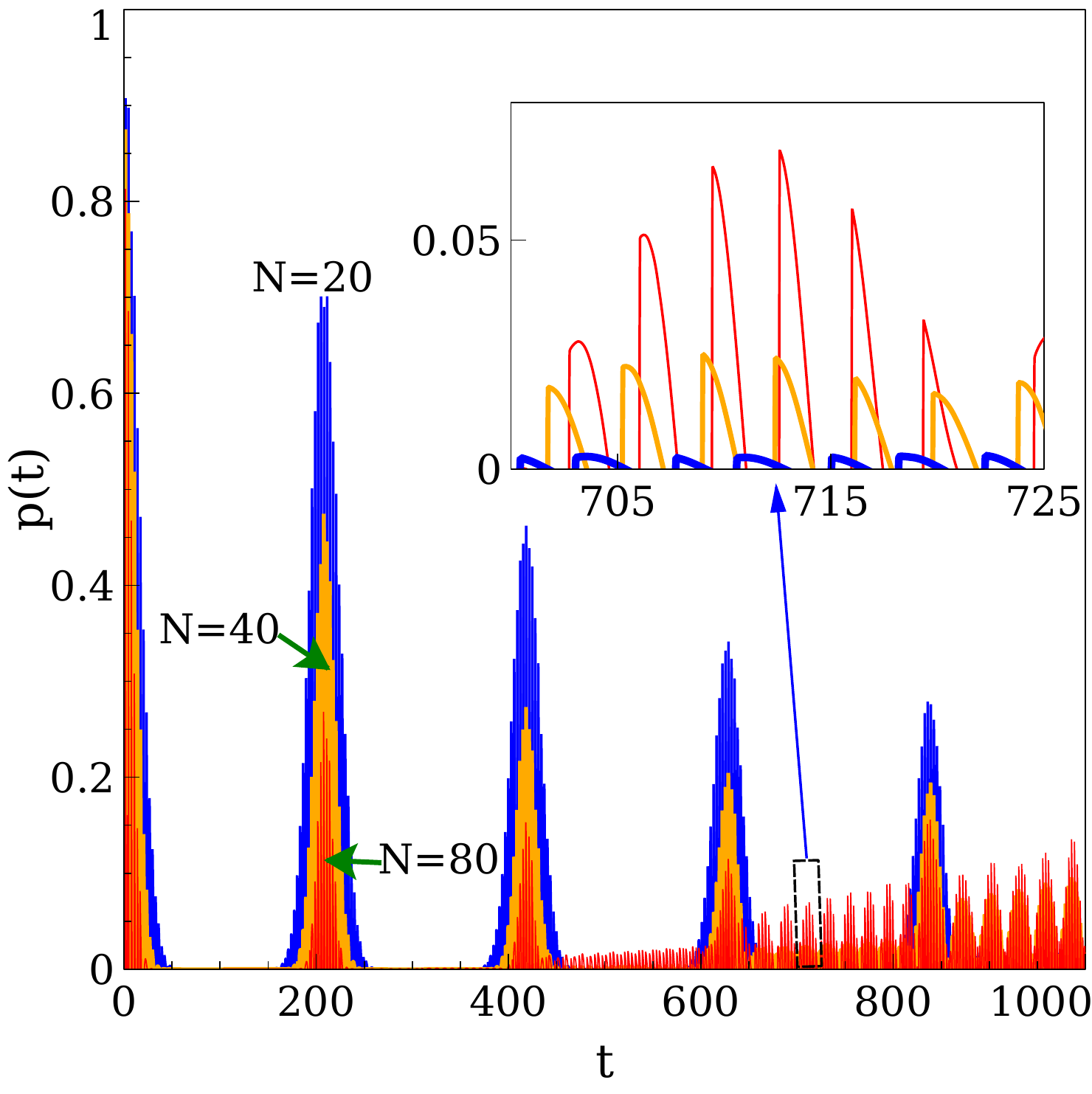}
	\caption{(Colour online) Variation of $p(t)$ with time $t$ for two density matrices $|\pm\ket=\frac{1}{\sqrt{2}}|0 \pm 1\ket$, for different number of bath spins $N$, and interaction strength $\alpha=0.03$. Magnified view of the rectangular region is shown in the inset.}
	\label{blp-N}
\end{figure}

Let us now investigate the aspect of non-Markovianity from another well known perspective, namely the distinguishability of two quantum states \cite{laine,breuer2}. Consider any distance measure $D(.)$ between two quantum states, following contraction property
\beq\label{sec1T}
D(\Phi[\rho^1],\Phi[\rho^2]) \leq D(\rho^1,\rho^2),
\eeq
where $\Phi[.]$ represents any CPTP map. Under any Markovian evolution, the time derivative of $D(.)$ will always be negative, owing to this contraction property. Therefore, non-monotonicity of these distances can be understood as a witness of the non-Markovian information feedback into the system. One such distance measure is the trace distance between quantum states \cite{ruskai}. Taking the trace distance between two states $D_T(\rho_1,\rho_2)=\frac{1}{2}||\rho_1-\rho_2||$  a quantity can be defined as
\beq\label{blp1}
p(t)=\frac{d}{dt}D_T(\Phi[\rho^1],\Phi[\rho^2]).
\eeq
Breuer-Laine-Piilo (BLP) proposed a  measure of non-Markovianity \cite{laine,breuer2} by summing over all the positive contributions of $p(t)$ and maximizing over the input states, which is given by
\beq\label{sec1W}
\varsigma=\max_{\rho^{1,2}}\int_{p(t)>0}p(t)dt.
\eeq 
It can readily be taken as a witness of non-Markovian information feedback into the system under any local decoherence channel. We find that for our specific quantum channel, the trace distance fidelity between two quantum states $\rho_1(t)$ and $\rho_2(t)$, at any arbitrary time after the action of the mentioned channel can be expressed as 
\beq\label{sec1U}
D(\Phi[\rho^1],\Phi[\rho^2]) = \sqrt{a^2(A(t)-B(t))^2+|b|^2|C(t)|^2},
\eeq 
with $a=\rho_{11}^1(0)-\rho_{11}^2(0)$ and $b=\rho_{12}^1(0)-\rho_{12}^2(0)$. 
In Fig. \ref{blp-fig}, we plot the function $p(t)$ for the two states $|\pm\ket=\frac{1}{\sqrt{2}}(|0\ket\pm|1\ket)$. The time evolution of the same is plotted in Fig. \ref{blp-N}, but for the case of increasing number of bath particles $N$. Note that calculating the maximized measure defined in Eq. (\ref{sec1W}), requires optimization over $a$ and $b$, which is difficult in general. However, consideration of two specific states can demonstrate the non-Markovianity providing a lower bound of the measure.
The two measures of non-Markovianity based on divisibility of the map (RHP measure, $\eta$) and distinguishability of two states under the action of the map (BLP measure, $\varsigma$) respectively, that we discuss here, may not agree in general \cite{kossa1,sabrina}. If a map is divisible, the evolution is Markovian and so the RHP measure of non-Markovianty $\eta$ is zero. Consequently the BLP measure $\varsigma$ is also zero. But the converse is generally not true, i.e., there exist some non-Markovian domain that are ``bound" in terms of BLP measure and hence not captured by it. The reason behind this is that the notion of complete positivity does not enter in BLP measure and hence the divisibility 
breaking cannot be fully captured by it \cite{kossa1}. In this work we also consider the BLP measure of non-Markovianity to study whether the non-Markovian feature of our proposed master equation can be captured by BLP measure also. 
\section{Negative entropy production rate}
\label{III}
The irreversible or nonequilibrium entropy production and its rate are two fundamental concepts in the analysis of the nonequilibrium processes and the performance of thermodynamic devices
\cite{entropy1,entropy2,entropy3,entropy4,entropy5,entropy6}. The reduction of the nonequilibrium entropy production can significantly alter the performance of thermodynamic devices and thereby it is of utmost interest in various technological domains.
  The nonequilibrium entropy production rate is defined as
\beq\label{sec2a}
\sigma(t) = \frac{dS}{dt}+ \mathcal{J},
\eeq
where $S$ is the von-Neumann entropy of the system and $\mathcal{J}=\frac{1}{KT}\frac{dQ}{dt}=\frac{1}{KT}\mbox{Tr}[H(t)\Lambda[\rho(t)]]$ is the entropy flux of the system. 
It can also be expressed as the time derivative of the relative entropy of the state $\rho$ with respect to the thermal equilibrium state $\rho_{eq}$ \cite{spohn,alicki} 
\beq\label{sec2b}
\sigma(t) = -\frac{d}{dt}S(\rho||\rho_{eq}), 
\eeq
where, $S\left(\rho||\tau\right)= -S(\rho)- \mbox{Tr}(\rho \ln \tau)$.
According to the Spohn’s theorem \cite{spohn} the nonequilibrium entropy production rate $\sigma$ is always non-negative. The Spohn's theorem is another statement of the second law of thermodynamics dictating the arrow of time. However, its validity essentially depends on the Markov approximation \cite{breuer2}. Under the non-Markovian dynamics $\sigma$ can be negative \cite{erez,gordon}. Therefore, the non-Markovianity of the dynamics is a thermodynamic resource providing partial reversibility of work and entropy. In addition, as negative $\sigma(t)$ is a prominent signature of the non-markovianity and hence it can be used to detect and quantify the non-Markovianity.
Since, for the specific system considered here, the absorption and the dissipation rates are equal due to the infinite temperature of the bath, the net heat flow $\frac{dQ}{dt}$ is always zero. Therefore, for this specific model, we have 
\beq\label{sec22}
\sigma(t) = \frac{dS}{dt}.
\eeq
It is worth mentioning that under the action of the unital channel von-Neumann entropy of a system always increases in Markovian dynamics, as it is also a doubly stochastic map. Since the given channel is unital, the negative $\frac{dS}{dt}$ also ensures the deviation from Markovianty. Note that the rate of change of entropy is given as 
\beq\label{sec2c}
\frac{dS}{dt} = -\frac{d}{dt}\left(\mbox{Tr}[\rho(t)\ln\rho(t)]\right)=-\mbox{Tr}[\ln\rho(t)\Lambda[\rho(t)]].
\eeq
Here $\Lambda[.]$ represents a general quantum evolution of the form
\beq\label{sec2d}
\begin{array}{ll}
\Lambda[\rho(t)]=-\frac{i}{\hbar}[\rho(t),H_S(t)]~~+\\
~~~~~~~~~~\sum_j \Gamma_j(t)\left[ V_j\rho(t)V_j^{\dagger}-\frac{1}{2}\{V_j^{\dagger}V_j,\rho(t)\} \right].
\end{array}
\eeq
If the Lindblad operators $\{V_j\}$ are Hermitian, then Eq. \eqref{sec2c} reads as 
\beq\label{sec2e}
\begin{array}{ll}
\frac{dS}{dt} = \frac{1}{2}\sum_{jkl}\Gamma_j(t) \left(\lambda_k(t)-\lambda_l(t)\right)~\times\\
~~~~~~~~~~~~[\ln\lambda_k(t)-\ln\lambda_l(t)]|\bra\lambda_k(t)|V_j|\lambda_l(t)\ket)|^2.
\end{array}
\eeq
where we take the spectral decomposition of the density matrix $\rho(t)=\sum_i \lambda_i(t)|\lambda_i(t)\ket\bra\lambda_i(t)|$. The above equation also implies that $\frac{dS}{dt}$ is non-negative if the relaxation rates $\{ \Gamma_j(t)\}$ are non-negative. However, $\frac{dS}{dt}$ can be negative if one or more of the relaxation rates $\{\Gamma_j(t)\}$ are negative, i.e, in the non-Markovian domain.
For the dynamics considered here, $\sigma(t)$ can be expressed as
\beq\label{sec2h}
\sigma(t) = \frac{1}{2}\ln\left(\frac{1-x}{1+x}\right)\frac{dx}{dt},
\eeq
 where $x=\sqrt{(\rho_{11}(t)-\rho_{22}(t))^2+4|\rho_{12}(t)|^2}$. We plot the nonequilibrium entropy production rate $\sigma(t)$ starting from the pure initial state $|1\ket$ in Fig. \ref{purity-entropy-fig}, which clearly shows that $\sigma(t)$ becomes negative whenever   $\Gamma_{dis}(t)$ becomes negative. It has been shown in Ref. \cite{erez} that for a diagonal qubit state,  $\sigma$ can be negative only when the non-Markovian dynamics drives the system away from its thermal equilibrium. The example considered here completely agrees with this fact.


 From Eq. (\ref{sec2e}) it is quite evident that the time rate of change of the entropy can be negative, only when the divisibility of the dynamical map breaks down. Therefore, a witness of non-Markovianity can be constructed from the negative entropy production rate for unital channels as follows
\beq\label{sec2j}
\varphi = \max_{\rho_{in}}\int_{\kappa(t)>0}\kappa(t)dt, 
\eeq
where $\kappa(t) = -\frac{dS}{dt}$.
%
%
Measure of the non-Markovianity based on the entropy production rate has been considered before for unital dynamical maps\cite{salimi}.
 

\subsection{Rate of change of purity: Detection of non-Markovianity}

\begin{figure}[htb]
	\includegraphics[width=7cm, height=6cm] {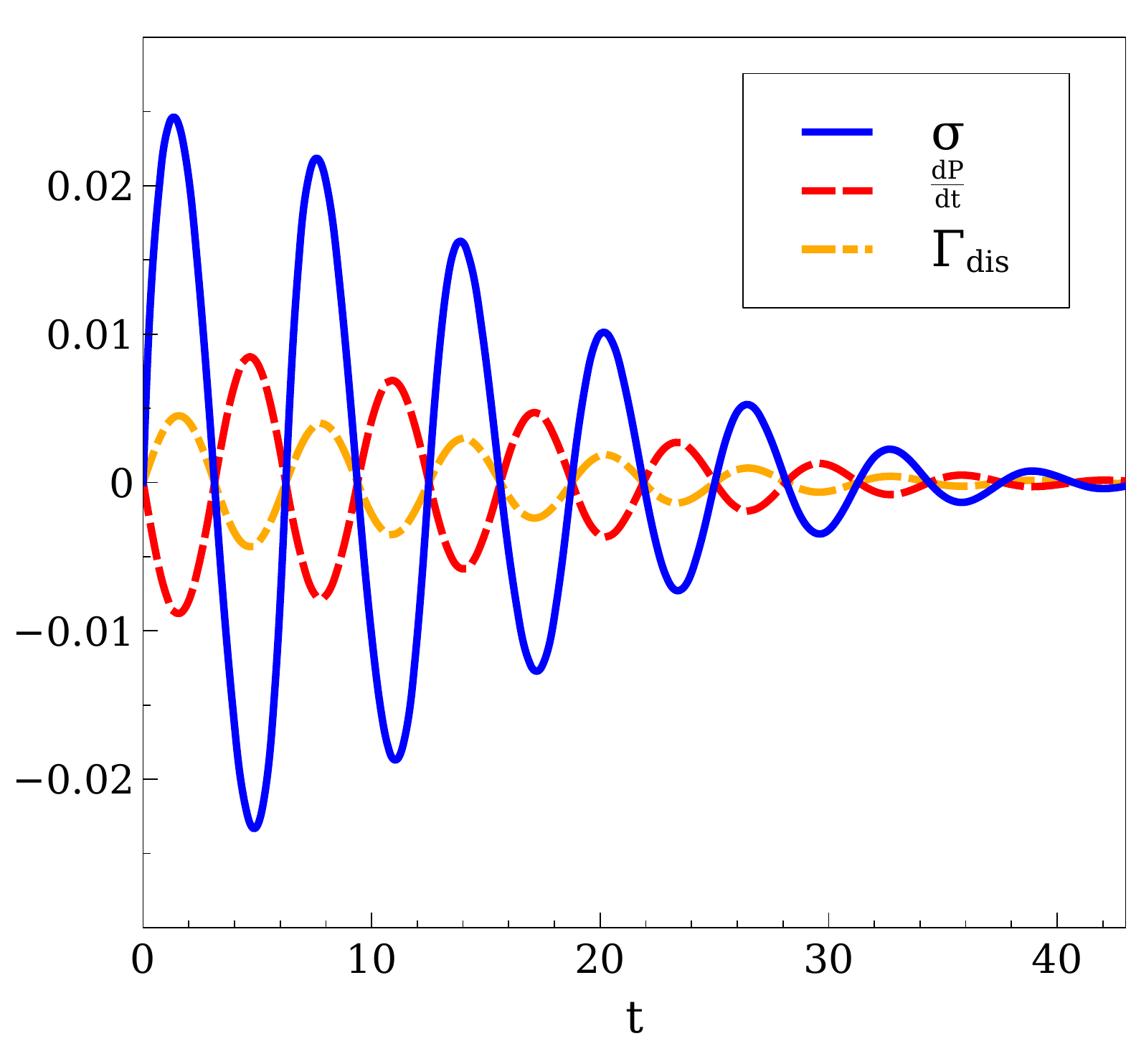}
	\caption{(Colour online) Variation of rate of change of the irreversible entropy production $\sigma(t)$, rate of change of the purity $\frac{dP}{dt}$, and $\Gamma_{dis}$ with time $t$ for the initial state $|1\ket$ with the interaction strength $\alpha=0.03$ and number of bath spins $N=20$. It is evident that $\sigma(t)$ and $\frac{dP}{dt}$ are negative and positive respectively, whenever $\Gamma_{dis}$ is negative. This implies that the non-Markovian information backflow revives purity of the state and causes negative irreversible entropy production rate.}
	\label{purity-entropy-fig}
\end{figure}

Let us investigate the non-Markovian behavior by the rate of change of the purity of the central qubit. If the Lindblad operators $\{V_j\}$ in Eq. \eqref{sec2d} are Hermitian then the rate of change of the purity $P(=\mbox{Tr} \rho^2)$, of the central qubit can be given as
\beq\label{depol2}
\frac{dP}{dt}=2 \mbox{Tr}[\rho(t)\Lambda[\rho(t)]]= -\sum_{i}\Gamma_{i}(t)Q_{i}(t),
\eeq
where $Q_{i}(t)=||\left[ V_{i},\rho(t)\right]||_{HS}^2$. The abbreviation in the subscript stands for Hilbert-Schmidt norm ($||X||_{HS}=\sqrt{Tr(X^{\dagger}X)}$). As $\{Q_{i}(t)\}$ are always positive, the positive rate of change of purity can only occur for the negativity of one or more of $\{ \Gamma_{i}(t)\}$ which corresponds to the divisibility breaking of the dynamical map. Note that the dynamics considered here can be expressed as a master equation with the Pauli matrices being the Lindblad operators (see Eq. \eqref{sec1Ia}) and the relaxation rates given as $\Gamma_x(t)=\Gamma_y(t)=\Gamma_{dis}(t)/2,~\Gamma_z(t)=\Gamma_{deph}(t)$. Since the Pauli matrices are Harmitian operators and thereby positive rate of change of purity of the central spin clearly signifies the non-Markovianity of the dynamical map. It is also worth mentioning that when the Lindblad operators $\{V_j\}$ are Hermitian or in other words when they represent observables, then $Q_{i}(t)=||\left[ V_{i},\rho(t)\right]||_{HS}^2$, measures 
the quantumness \cite{facchi,samya} of the state $\rho(t)$. Therefore, Eq. \eqref{depol2} implies that the more is the quantumness of the state the more it is sensitive to the environment.
  After a little algebra, we find that the rate of change of purity for the initial central qubit state $|1\ket$, is given as
\beq\label{depol3}
\frac{dP}{dt} =  [A(t)-B(t)]\frac{d}{dt}[A(t)-B(t)].
\eeq
We plot the rate change of the purity with time in Fig. \ref{purity-entropy-fig}.
 From Fig. \ref{purity-entropy-fig} it can be seen that the positive rate of change of purity occurs periodically, whenever the relaxation rate $\Gamma_{dis}(t)$ is negative. Since we are taking a initial diagonal state in the computational basis, there is no effect of the dephasing channel on the central qubit. For a qubit  system, its eigenvalues have the form
$\lambda = \frac{(1\pm z)}{2}$, where $0\leq z \leq1$, and hence, the entropy of a qubit system is a monotonically decreasing function of the purity of the qubit. Therefore, signs of the rates of change of purity and entropy (See Fig. \ref{purity-entropy-fig}) are opposite.
%
%
%

%
%

Nowadays with advanced experimental techniques, the purity of  a quantum system can be directly measured \cite{zoller,ekert,bovino}. Hence, the non-Markovian revival of purity can be experimentally verified to demonstrate the non-Markovianity and the negative nonequilibrium entropy production rate in the laboratory.

\section{Conclusion}
\label{IV}
 To summarize, we have considered dynamics of a central spin-half particle which is interacting with a bath consisting of completely unpolarized, non interacting spin half particles. An exact canonical Lindblad type master equation has been derived for the central spin system. The dynamics of the system exhibits  non-Markovian features which have been characterized and quantified  by divisibility breaking (the RHP measure of non-Markovianity ) as well as 
monotonicity breaking of the trace distance fidelity (the BLP measure of non-Markovianity) conditions. The Kraus operators for the dynamical evolution is also derived.

The nonequilibrium entropy production rate has been investigated. Negative entropy production rate implies the non-Markovianity of the dynamics, though the reverse does not hold true. The dynamics of the central spin considered here shows that for a specific initial state the non-Markovianity of the dynamics is always associated with negative entropy production rate. Moreover, it has also been shown that in this dynamics, the non-Markovianity is always accompanied by the increase of the purity of the central spin when the same initial state has been chosen. As the purity is a measurable quantity, the exact canonical Lindblad type master equation of the central spin, derived in this article, could be of paramount importance to investigate the non-Markovian features and negative entropy production rate in the laboratory. The scheme used here to derive the canonical master equation has been proven to be fruitful to explore the strong coupling regime where the system-bath separability breaks down, which gives 
the present study a practical importance to unravel the far reaching impacts of the non-Markovian dynamics in the strong coupling regime in various information theoretic and thermodynamic protocols and devices.

\section*{Acknowledgement}

Authors acknowledge M.J.W. Hall for his useful remarks and suggestions on the manuscript and financial support from the Department of Atomic Energy, Govt. of India.

\bibliographystyle{apsrev4-1}
\bibliography{spin-bath}

\end{document}